\documentstyle[sprocl]{article}

\def\Journal#1#2#3#4{{#1} {\bf #2}, #3 (#4)}

\def\NPB{{\em Nucl. Phys.} B}
\def\PLB{{\em Phys. Lett.}  B}

\def\al{\alpha}
\def\om{\omega}
\def\Om{\Omega}

\def\ca{{\cal{A}}}
\def\cd{{\cal{D}}}
\def\cg{{\cal{G}}}
\def\co{{\cal{O}}}
\def\shalf{\frac{1}{2}}
\def\be{\begin{equation}}
\def\ee{\end{equation}}
\def\bea{\begin{eqnarray}}
\def\eea{\end{eqnarray}}
\begin{document}

\title{EXERCISES IN EQUIVARIANT COHOMOLOGY AND TOPOLOGICAL THEORIES}

\author{ R. STORA }

\address{Laboratoire de Physique Th{\'e}orique ENSLAPP \footnote{URA 1436 du 
CNRS, associ{\'e}e {\`a} l'Ecole Normale Sup{\'e}rieure de Lyon et {\`a}
l'Universit{\'e} de Savoie.}, B.P. 110, \\ F-74941 Annecy-le-Vieux
Cedex, France
\\and\\ Theory Division, CERN, CH-1211, Geneva 23, Switzerland.}

\maketitle\abstracts{
Equivariant cohomology is suggested as an alternative algebraic
framework for the definition of topological field theories
constructed by E.~Witten circa 1988. It also enlightens the
classical Faddeev Popov gauge fixing procedure.}
   
\section{Introduction}\label{sec:I}

Before going into the subject of this talk, I would like to describe
some concrete exercises done by Claude and I which represent a very
small portion of the numerous discussions we had, mostly by exchange
of letters. We happened to be both guests of the CERN theory
division during the academic year 1972-1973.

The perturbative renormalization of gauge theories was still a hot
subject, and, whereas most of our colleagues considered the problem
as solved we were both still very innocent. I happened to be
scheduled for a set of lectures for the "Troisi{\`e}me cycle de la
Suisse Romande" in the spring 1973, on the subject "Models with
renormalizable Lagrangians: Perturbative approach to symmetry
breaking", and I decided to conclude those lectures with a summary
of the known constructions related to gauge theories, mostly at the
classical level, except for a heuristic derivation of the now
called \cite{1} Slavnov Taylor identities, taking seriously the
Faddeev Popov ghost and antighost as local fields. What had to be
done was indicated in A. Slavnov's preprint which I had remarked:
perform a gauge transformation of parameter $m^{-1} \bar{\xi}$
where $m$ is the Faddeev Popov operator and $\bar{\xi}$ the source
of the antighost field. That strange trick was due to E.S. Fradkin and I.V. Tyutin as
indicated in Slavnov's preprint. At the time, I was not aware of
J.C. Taylor's paper which came to my attention much later. Anyway,
Claude and I carried out that calculation whose result is reported
in the notes, with details in an appendix for which the authors (A.
Rouet and I) thank Claude Itzykson for generous help \cite{2}. It is
that form of the identity which, a few months later drew Carlo
Becchi and Alain Rouet's attention, leading them to the remark that
the gauge fixed Faddeed Popov action possesses a symmetry naturally
called the Slavnov symmetry. A year later, when the paper by E.S.
Fradkin and G.A. Vilkovisky on the quantization of canonical systems
with constraints came out, Claude and I had a conversation on the
telephone and we found we had both noticed that paper. I suggested
that the action they proposed possessed a Slavnov symmetry. A couple
of days later, Claude called me back and gave me the formula -at
least in the case of gauge constraints- which I immediately forgot.
When I met E.S. Fradkin in Moscow in the fall 1976, I told him about
Claude's finding, and there followed the first article by I.A.
Batalin and G.A. Vilkovisky who unfortunately thank me for
suggesting the problem, and do not mention Claude at all.

These are only two examples of the innumerable discussions we had on
physics and other things as well, mostly in writing, because life
did not make our trajectories intersect so often. The last long
series of discussions I had with him took place in Turku, Finland,
at the meeting of the spring 1991. Almost every evening, we were
ambulating around the big lawn in front of the dining room, trying
to reconstruct, at his request, the arguments which produce the
existence of 27 straight lines on an unruled third degree surface.
That was a prelude to his later work on enumerative geometry.

Generous, he was; intelligent he was; cultivated he was; we remain
deprived of patiently gathered wisdom, a rather rare item.

Returning to technicalities I will now try to describe a few facts
about the Lagrangian formulation of topological -more precisely
cohomological- field theories, constructed by E. Witten from 1988 on,
 in as much as they are relevant to our poor
understanding of gauge theories. That is to say I will insist on the
field theory aspects in particular, the distinction between fields
and observables, even though a host of beautiful results and
conjectures have been obtained otherwise.

Equivariant cohomology is roughly forty five years old, and yet,
does not belong to most theoretical physicists' current mathematical
equipment. The easy parts, namely, definitions, terminology,
elementary properties are described in the appendix whose content is
freely used throughout the text.

Section~\ref{sec:II} is devoted to a reminder on dynamical gauge
theories and a formal description of the Faddeev Popov gauge fixing procedure in
terms of notions belonging to the theory of foliations \cite{3}.

Section~\ref{sec:III} describes some aspects of "cohomological"
topological theories with emphasis on some of the features which distinguish
them from dynamical theories at the algebraic level provided by the
Lagrangian descriptions.

\section{Formal aspects of dynamical gauge theories}\label{sec:II}

Here are a few considerations on formal aspects of the Faddeev Popov
gauge fixing procedure which allowed to handle, thanks to the very
strong consequences of locality, the ultraviolet difficulties found
in the perturbative treatment of theories of the Yang Mills type.
This can be found in most textbooks and usually proceeds via
factoring out of the relevant functional integral the infinite
volume of the gauge group produced by the gauge invariance of the
functional measure. There is a more satisfactory strategy sketched
in J.~Zinn Justin's book \cite{4} which avoids this unpleasant step,
and fits more closely mathematical constructions now classical in
the theory of foliations \cite{3}.

The set up is as follows:

$M_4$ is a smooth space time manifold, which one may choose compact
without boundary, in euclidean field theory. $P(M,G)$ is a principal
$G$ bundle over $M_4$, $\bigcup\limits_i {(U_i \times G)}$
 modulo glueing maps above $U_i \cap U_j$, where $\{U_i\}$ is an open
covering of $M$). $G$ is a compact Lie group referred to as the
structure group. $\ca$ is the set of principal connections $a$ on
$P(M,G)$ (Yang Mills fields). On $M_4$
\be
a_M = \sum_\al a^\al_\mu (x) dx^\mu e_\al \ \ \ \ \ e_\al:
\mbox{basis of Lie} \ G
\label{eq:1}
\ee

On $P(M,G)$, locally,
\be
a=g^{-1} a_M \ g + g^{-1} dg \ \ \ \ (x,g) \ \mbox{local coordinates
in} \ U \times G
\label{eq:2}
\ee
\be
F(a) = da + \shalf [a,a]
\label{eq:2bis}
\ee
is the curvature of $a$ (the field strength).

$\ca$ is acted upon by $\cg$, the gauge group, i.e. the group of
vertical automorphisms of $P(M,G)$ ("gauge transformations"). Upon
suitable restrictions, $\ca$ is a principal $\cg$ bundle over $\ca 
/ \cg$, the set of gauge orbits.

Dynamical gauge theories are models in which the fields are the
$a$'s (and, possibly matter fields), and the observables are gauge
invariant functions of the $a$'s (or functions on $\ca / \cg$).

For historical as well as technical reasons related to locality, one
chooses models specified by a local gauge invariant action
\be
S_{YM} (a) = \frac{1}{4g^2} \int_{M_4} tr F \wedge  \ast F.
\label{eq:3}
\ee

Heuristically, one considers the $\cg$ invariant measure on $\ca$
\be
\Omega_{YM} = e^{-S_{YM}(a)} \underbrace{\wedge \ \delta a}_{\cd a}
\label{eq:4}
\ee

If $\{ X_\al \}$ denotes a basis of fundamental vertical vector
fields representing the action of Lie $\cg$ on $\ca$, one constructs the
Ruelle Sullivan \cite{5} current
\be
\Omega_{RS} = i(\mathop \Lambda \limits_\alpha X_\al)
\Omega_{YM} \label{eq:5}
\ee
which is closed and horizontal, therefore basic: (cf. Appendix A)
\bea
\delta \Omega_{RS} &=& 0 \nonumber \\
i(X_\al) \Omega_{RS} &=&0
\label{eq:6}
\eea
hence
\be
\ell(X_\al) \Omega_{RS} =0
\label{eq:7}
\ee

It follows in particular that $\Omega_{RS}$ is invariant under field
dependent gauge transformations.

Given a gauge invariant observable $\co(a)$, the question is to
integrate it against $\Omega_{RS}$, or rather to integrate its image
as a function on $\ca / \cg$ against the image of $\Omega_{RS}$ as a
top form on $\ca / \cg$.

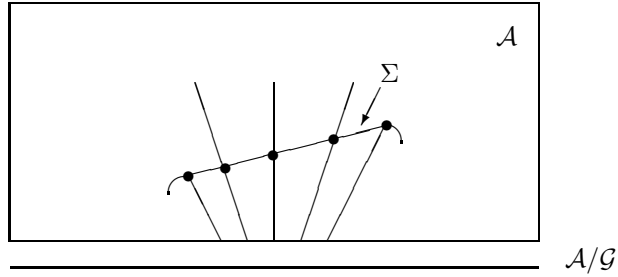
\begin{center}
\begin{picture}(200,120)(0,0)
\put(0,10){\framebox(200,90)}
\put(184,84){$\cal A$}
\put(0,0){\line(1,0){200}}
\put(210,0){${\cal A}/{\cal G}$}
\put(80,10){\line(-1,2){12}}
\put(90,10){\line(-1,3){20}}
\put(100,10){\line(0,1){60}}
\put(110,10){\line(1,3){20}}
\put(120,10){\line(1,2){22}}
\put(66,34.5){\line(4,1){75}}
\put(140,70){$\Sigma$}
\put(140,68){\vector(-1,-2){7}}

\put(65,32){$\bullet$}
\put(67,27.5){\oval(14,14)[tl]}
\put(79,35){$\bullet$}
\put(97,40){$\bullet$}

\put(120,46){$\bullet$}
\put(140,51){$\bullet$}
\put(141,47){\oval(14,14)[tr]}

\end{picture}
\end{center}
\indent

Choose a local section $\Sigma$ (transverse to the fibers) with
local equations
\be
g(a) =0
\label{eq:8}
\ee
and corresponding local coordinates $\dot{a}$ so that a local
parametrization of $\ca$ is given by
\be
a=\dot{a}^g
\label{eq:9}
\ee
i.e. all $a$'s are, locally gauge transforms of points on the chosen
transversal manifold. 

One can represent the transverse measure associated with the chosen
section as follows:
\bea
<\co>_\Omega = \int_\Sigma \co(\dot{a}) \Omega_{RS|\Sigma}
&=&\int_\Sigma \co(\dot{a}) \Omega_{RS|\Sigma}
{\underbrace{\int_{fiber}
\delta(g) \wedge \delta g}_{=1}}_! \nonumber \\
&=& \int \co(a) \Omega_{RS} \ \delta(g) det m (\wedge g^{-1}
\delta g)
\label{eq:10}
\eea
where the volume $\wedge g^{-1} \delta g$ is chosen so that
\be
i(\wedge_{\dot{\al}} X_{\dot{\al}}) (\wedge g^{-1} \delta g) =1
\label{eq:11}
\ee
and $m$ is given by
\be
m= \frac{\delta g}{\delta a} D_a
\label{eq:12}
\ee

Thus
\be
\Omega_{RS} (\wedge g^{-1} \delta g) = \Omega_{YM}
\label{eq:13}
\ee
and the result follows:
\be
<\co>_\Omega = \int \co(a) \  \delta (g) \ \det m \ \Omega_{YM}
\label{eq:14}
\ee

This, of course only holds if $\co(a)$ has its support inside the
chosen chart. By construction, the result is independent of the
choice of a local section, two local sections differing by a field
dependent gauge transformation.

The final outcome is to replace $\Omega_{YM}$ by 
\be
\Omega_{YM \Phi \Pi} = \Omega_{YM} \ \Omega_{\Phi \Pi}
\label{eq:15}
\ee
where
\be
\Omega_{\Phi \Pi} = \int \cd \bar{\omega} \cd \omega \cd b
\ e^{i<b,g(a)> + <\bar{\omega}, m \omega>}
\label{eq:16}
\ee
where we have used the Stueckelberg Nakanishi Lautrup Lagrange
multiplier $b$, the Faddeev Popov fermionic ghost $\omega$, the
Faddeev Popov fermionic Lagrange multiplier (antighost) $\bar{\om}$.
The modern reading of the exercise done with Claude is that not
only $\Om_{YM \Phi \Pi}$ is invariant under the operation $s$
\bea
sa &=& - \cd_a \om \nonumber \\
s\om &=& - \shalf [\om, \om] \ \ \ \ s^2 =0 \nonumber \\
s \bar{\om} &=& -ib \nonumber \\
sb &=& 0
\label{eq:17}
\eea
but, thanks to the introduction of the $b$-field,
\be
i<b,g> + <\bar{\om} , m\om> = s \left( -<\bar{\om} ,g> \right)
\label{eq:18}
\ee

This allows to discuss perturbative renormalization using all the
power of locality. The useful part involves the local cohomology of
Lie $\cg$ in terms of which the observables can be defined and which
also classifies obstructions to gauge invariance due to quantum
deformations (i.e. anomalies).

We shall see in the next section that the cohomology involved in
topological theories is different !

Of course the above discussion is local over orbit space, and a
constructive procedure to glue the charts is missing. This is the
Gribov problem.

\section{Cohomological Theories}\label{sec:III}

E. Witten's 1988 paper \cite{6} contains several things. First,
invoking "twisted $N=2$ supersymmetry" E.~Witten gets an action
$S(a, \psi, \varphi;...)$ where $\psi$ resp $\varphi$ is a 1 resp 0
 form with values in Lie $G$ and the dots represent a collection of
 Lagrange multiplier fields. Then it is observed that
 \be
 QS=0
 \label{eq:19}
 \ee
 with
 \be
 \begin{array}{cc}
 Qa=\psi & \ \ \ \ \mbox{infinitesimal} \\
 Q \psi = D_a \varphi & \ \ \ \ Q^2 = \mbox{gauge transformation} \\
 Q\varphi =0 & \ \ \ \ \mbox{of parameter} \ \varphi
 \end{array}
 \label{eq:20}
 \ee
 
 Furthermore there is an identity of the form
 \be
 \int tr F \wedge F = S - Q \chi(a,\psi, \varphi;...)
 \label{eq:21}
 \ee
 where $\chi$ is gauge invariant.
 
 The observables are classified according to the gauge invariant
 cohomology of $Q$, with the example
 \bea
 Q \ tr \ F \wedge F &=& -d \ tr \ 2F\psi \nonumber \\
 Q \ tr \ 2F \psi &=& -d \ tr \ (\psi \wedge \psi + 2F\varphi)
 \nonumber  \\ 
 Q \ tr (\psi_\wedge \psi + 2F \psi)  &=& -d (2 \psi \varphi)
\nonumber \\
Q \ tr \ 2 \psi \varphi &=& -d \ tr \ \varphi^2 \nonumber \\
Q \ tr \ \varphi^2 &=& 0
\label{eq:22}
\eea

It follows that integrating the polynomials exhibited in these
descent equations over cycles of the correct dimensions yields (non
trivial !) elements of the cohomology of $Q$ whose correlation
functions are conjectured to reproduce Donaldson's polynomials.

Very soon after the appearance of E.~Witten's article, L.~Baulieu
and I.M.~Singer \cite{7} remarked that Eq.(\ref{eq:21}) can be
rewritten as
\be 
S= \int tr \ F \wedge F+Q\chi (a, \psi, \varphi; ...)
\label{eq:23}
\ee
so that this action looks like the gauge fixing of a topological
invariant. Furthermore, at the expense of introducing a Faddeev
Popov ghost $\om$, $Q$ can be replaced by $s$:
\bea
sa &=& \psi - \cd_a \om \nonumber \\
s\psi &=& - D_a \Om + [\psi, \om] \ \ \ \ s^2 \equiv 0 \nonumber \\
s\om &=& \Om - \shalf [\om, \om] \nonumber \\
s \Om &=& -[ \om, \Om] \nonumber \\
\label{eq:24}
\eea

(For homogeneity in the notations, we have replaced $\varphi$ by
$\Om$).

This has however a defect, namely, $s$ has no cohomology and
therefore is not adequate to describe the physics of the model.

Inspired by an article by J.~Horne \cite{8}, devoted to a
supersymmetric formulation of this model, S.~Ouvry, R.~S. and
P.~van~Baal \cite{9} solved that difficulty by phrasing J.~Horne's
observation as follows: $S$ and $\chi$ are not only gauge invariant
but also are independent of $\om$ !

In other words they are invariant under
\be
\begin{array}{lll}
I(\lambda), L(\lambda) , & \lambda \in \mbox{Lie} \  \cg &  \\
I(\lambda) \om = \lambda & I(\lambda) \ \mbox{other} =0  &\\
L(\lambda) \om = [\lambda, \om] & L(\lambda)  \ \mbox{other =} &
\mbox{infinitesimal  gauge}\\
& & \mbox{transformation of parameter} \ \lambda 
\label{eq:25}
\end{array}
\ee
and, one can verify that
\be
L(\lambda) = [I(\lambda), s]_+
\label{eq:26}
\ee

The cohomology that defines the physics of the model is the basic
cohomology of $s$ for the operation $\{I(\lambda), L(\lambda) \}$.
This is not empty and co{\"\i}ncides with that of $Q$. Looking into
that direction was suggested during a seminar by P.{\S} Braam at the CERN
theory division in the spring 1988. There it was stated that the
subject was the equivariant cohomology of $\ca$ (restricted to
$F=\ast F$). Further geometrical interpretations of $\psi \om \Om$
were given by L.~Baulieu and I.M.~Singer \cite{7} and the general set
up was precisely phrased in terms of equivariant cohomology by
J.~Kalkman \cite{10} who developed the algebraic equipment further.
Two general types of equivariant cohomology classes are involved in
the present models:

- Matha{\"\i}~Quillen \cite{11} representatives of Thom class of
vector bundles (Gaussian deformations of covariant $\delta$
functions). Those occur in the action.

- Equivariant characteristic classes of vector bundles. They are
expressed in terms of an arbitrary invariant connection\cite{12}.
They provide the known topological observables. In the case where
the manifold to be quotiented is a principal bundle, Cartan's
"theorem 3"\cite{13} transforms equivariant cohomology classes into
basic cohomology classes, by the substitution $\om \rightarrow
\tilde{\om}, \Om \rightarrow \tilde{\Om}$, where $\tilde{\om}$ is a
connection and $\tilde{\Om}$ its curvature. It is expressible in
terms of another identity in which integral representation of both
bosonic and fermionic $\delta$ functions provides other terms in the
action:
\be
\int \cd \om \cd \Om \ \delta (\om - \tilde{\om}) \  \delta (\Om -
\tilde{\Om}) =1
\label{eq:27}
\ee

This can only be understood if $\om$ is introduced, although it does
not always appear in the action.

We shall now illustrate these general recipes in the case of
topological Yang Mills theories $(YM^{top}_4)$.

The observables are constructed as universal cohomology classes of
$\ca / \cg$ as follows: consider the $G$ bundle $P(M,G) \times \ca$
and, on it, the $\cg$ invariant $G$ connection $a$ (a zero form on
$\ca$, a one form on $P(M,G)$).

The equivariant curvature of $a$, in the intermediate scheme (see
appendix A) is
\be
R^{eq.}_{int} = F(a) + \psi + \Om
\label{eq:27b}
\ee
with 
\be
\psi = \delta a.
\label{eq:28}
\ee

In the Weil scheme, we are interested in
\be
R^{eq.}_w = F(a) + \psi + \Om
\label{eq:29}
\ee
with
\be
\psi = \delta a + \cd_a \om.
\label{eq:30}
\ee

This is the object first considered by L.~Baulieu,
I.M.~Singer \cite{7}.

The equivariant characteristic class $tr (R^{eq.}_w)^2$
fulfills
\be
(d+\delta) \ tr (R^{eq.}_w)^2 =0
\label{eq:31}
\ee
which provides the descent equations (Eq.\ref{eq:22}). Replacing
$\om$ by $\tilde{\om}, \Om$ by $\tilde{\Om}$, where $\tilde{\om}$ is
a $\cg$ connection on $\ca$, provides a basic form on $P(M,G) \times
\ca$.

One may choose \cite{7},\cite{11}
\be
\tilde{\om} = - D^*_a \frac{1}{D^*_a D_a} \delta a
\label{eq:32}
\ee
provided reducible connections are excluded.

Let now $\co_i (a, \psi, \om, \Om)$ be equivariant classes of
$\ca$ obtained by integration over cycles in $M$ with the proper
dimension. We want to find an integral representation in terms of
fields of the form on $\ca / \cg$ corresponding to a basic form
$\co = \prod_i \co_i$ and, in the case of a form of maximal
degree ("top form") of its integral.

Let $\tilde{a}$ be coordinates of a local section $\Sigma$
\be
g(\tilde{a}) \equiv 0 \ \ \frac{\delta g}{\delta a} \delta \tilde{a}
\equiv \frac{\delta g}{\delta a} (\tilde{\psi} - D_{\tilde{a}}
\tilde{\om}) \equiv 0
\label{eq:33}
\ee

We have
\be
\co (a, \psi, \tilde{\om}, \tilde{\Om})_{| \Sigma} = \co
\left( \tilde{a}, \delta \tilde{a} + D_{\tilde{a}} \tilde{\om}_{|
\Sigma}, \tilde{\om}_{| \Sigma}, \tilde{\Om}_{| \Sigma} \right)
\label{eq:34}
\ee

This defines a cohomology class on $\ca / \cg$, independently of the
choice of $\Sigma$, because of the basicity of $\co$. The
expression at hand can be expressed through the introduction of a
collection of $\delta$-functions.

First, in the case of $YM^{top}_4$, one has to restrict to
$F=\ast F$, which goes through a $\delta$ function or a smeared
gaussian thereof according to the Matha{\"\i} Quillen formula (cf.
Ref.\cite{11} and appendix A).

The replacement $\om \rightarrow \tilde {\om} \ \Om \rightarrow 
\tilde{\Om}$ can be carried out using the $\delta$ functions of
Eq.(\ref{eq:27}):
\bea
&&\int \delta(\om - \tilde{\om}) \delta(\Om - \tilde{\Om}) \cd \om
\cd \Om \nonumber \\
&&= \int \cd \bar{\om} \cd \bar{\Om} \cd \om \cd \Om
\ e^{(s+\delta)(\bar{\Om}(\om - \tilde{\om}))}
\label{eq:35}
\eea
where $s$ is extended to
\bea
s \bar{\Om} &=& \bar{\om} - [ \om, \bar{\Om}] \nonumber \\
s \bar{\om} &=& [ \Om, \bar{\Om}] - [\om, \bar{\om}]
\label{eq:36}
\eea

If $\tilde{\om}$ is the solution of a local equation e.g.
\be
D_a^{\ast} \tilde{\Psi} = D^\ast_a (\delta a + D_a \tilde{\om})
\label{eq:37}
\ee
this can be rewritten, thanks to the cancellation of determinants,
as:
\be
\int
\cd \om \cd \Om \cd \bar{\om} \cd \bar{\Om} \ e^{s(\bar{\Om} D^\ast
\Psi)}
\label{eq:38}
\ee

Other local choices can be made, e.g. the flat connection determined
by the local section $\Sigma$ \cite{14}, but, in this case, a change
of local section produces a change of representative in
the cohomology class under consideration due to the associated
change of connection.

Finally, the restriction to $\Sigma$ goes via the insertion of the
$\delta$ function identity
\be
\int \delta (a-\tilde{a}) \delta (\psi -
\tilde{\psi}) \cd a \cd \psi =1
\label{eq:39}
\ee
This can be rewritten as
\be
\int \cd a \cd \psi \int \cd \bar{\al} \cd \bar{\psi}
\ e^{(s+\delta)(\bar{\psi}(a-\tilde{a}))} =1
\label{eq:40}
\ee
with
\bea
s \bar{\psi} &=& \bar{\al} - [\om, \bar{\psi}] \nonumber \\
s \bar{\al} &=& [\Om, \bar{\psi}] - [ \om, \bar{\al}]
\label{eq:41}
\eea

Integrating over all $a$'s and $\Psi$'s yields a field theory
representation of forms on orbit space, as advocated in
ref.\cite{14}. Integrating over the superfiber (the tangent bundle
of a fiber with Grassmann variables on the vectorial part) yields a
formal field theory representation of the integral over orbit space
of a basic top form. In terms of the local equations
Eq.(\ref{eq:33}), this can be rewritten as
\be
\int \cd a \cd \psi \int \cd \bar{\al} \cd \bar{\psi}
\ e^{s(\bar{\gamma} g(a))} =1
\label{eq:42}
\ee
with
\bea
s \bar{\gamma} &=& \beta + \om \cdot \bar{\gamma} \nonumber \\
s \beta &=& - \Om \cdot \bar{\gamma} + \om . \beta
\label{eq:43}
\eea
where the dot denotes the action of $\cg$ on the bundle over $\ca$
of which $g$ is a section.

If $\co$ is a top form, integration transforms the integration
over the fiber, in Eqs (\ref{eq:40}, \ref{eq:41}) into integration
over $\ca$, after localizing $\co$ inside the domain of $\Sigma$.
The result is then a functional integral of the exponential of an
action of the form $s \chi$. If this representation involves
ultraviolet problems one may conjecture that, besides the necessity
to include in $\chi$ all terms consistent with power counting the
gauge fixing term in Eq.(\ref{eq:42}) has to be written in the form
$sW\chi$ where $W$ is another operation which anticommutes with
$s$ and involves a Faddeev Popov ghost field, its graded partner, and
the corresponding antighosts. This however is still waiting for
confirmation.

In support of the relevance of these constructions, one may give a
few examples:

i) The equivariant curvature Eq.(\ref{eq:29}),(\ref{eq:31})
precisely yields the observables constructed by E.~Witten via the
interpretation given by L.~Baulieu, I.M.~Singer. The same method
yields the observables constructed by C.~Becchi, R.~Collina,
C{\'E}~Imbimbo \cite{14} in the case of 2-d topological gravity (see
also L.~Baulieu, I.M.~Singer \cite{7}).

ii) Recent work by M.~Kato \cite{15} and collaborators remarking the
equivalence of some pairs of topological conformal models through
similarity transformations of the form $e^R$ is interpretable by
$R=i_M (\om)$, in J.~Kalkman's language~\cite{10}.

iii) The identification in topological actions of terms which fix a
choice of connection is an additional piece of evidence \cite{6},
\cite{14}.

\section{Conclusion}

The formalism of equivariant cohomology provides an elegant
algebraic set up for topological theories of the cohomological type.
Its relationship with $N=2$ supersymmetry via twisting is still
mysterious and may still require some refinements before it provides
some principle of analytic continuation. At the moment, it is still
a question whether topological theories can be treated as field
theories according to strict principles \cite{14} or whether the
formal integral representations they provide can at best suggest
mathematical conjectures to be mathematically proved or disproved.

\section*{Acknowledgments}
I wish to thank C.~Becchi and C.~Imbimbo for numerous discussions
about their work on 2d topological gravity. I also wish to thank
R.~Zucchini for discussions about his recent work.

\section*{Appendix A}

 {\bf Equivariant Cohomology}
 
 \underline{Example 1.}
 
 $M$ is a smooth manifold with a smooth action of a connected Lie
 group $\cg; \Om^\ast (M)$ is the exterior algebra of differential
 forms on $M, d_M$ the exterior differential; $\lambda \in$ Lie
 $\cg$ is represented by a vector field $\underline{\lambda} \in
 \mbox{Vect} M. i_M (\lambda) = i(\underline{\lambda})$ operates on
 $\Om^\ast (M)$ by contraction with $\underline{\lambda}$; the Lie
 derivative is defined by 
 \be
 \ell_M(\lambda) = \ell(\underline{\lambda}) =
 [i(\underline{\lambda}), d_M  ]_+
 \label{eq:A1}
 \ee
 
 One has
 \bea
 {[}i_M(\lambda), i_M(\lambda ')]_+ &=& 0 \nonumber \\
 {[}\ell_M(\lambda), i_M(\lambda ')]_- &=& i_M([\lambda, \lambda '])
 \nonumber \\
 {[}\ell_M(\lambda), \ell_M(\lambda ')]_- &=& \ell_M ([\lambda,
 \lambda  '])  \label{eq:A2}
\eea

Forms $\om \in \Om^\ast (M)$ such that
\be
i_M(\lambda) \om =0 \ \ \ \forall \lambda \in \mbox{Lie} \  \cg
\label{eq:A3}
\ee
are called horizontal.

Forms $\om \in \Om^\ast (M)$ such that
\be
\ell_M(\lambda)\om =0 \ \ \ \forall \lambda \in \mbox{Lie} \  \cg
\label{eq:A4}
\ee
are called invariant.

Forms which are both horizontal and invariant are called basic.

The basic de Rham cohomology is the cohomology of $d_M$ restricted
to basic forms.

\underline{Generalization.}

$E$ is a graded commutative differential algebra with differential
$d_E$ and two sets of graded derivations $i_E(\lambda)$ (of
grading -1) $\ell_E(\lambda)$ (of grading 0) fulfilling
Eq.(\ref{eq:A2}), with $M$ replaced by $E$. The notions of
horizontal and invariant elements similarly generalize as well as
that  of basic cohomology.

Example 2: The Weil algebra of $\cg: W(\cg)$.
\be
W(\cg) = \wedge (\mbox{Lie} \ \cg)^\ast \otimes S\left( (\mbox{Lie}
\ \cg )^\ast \right)
\label{eq:A5}
\ee
whose factors are generated by $\om$, of grading $1, \Om$ of grading
2, with values in Lie $\cg$. We define the differential $d_w$ by
\bea
d_W \ \om &=& \Om - \shalf [\om, \om]
\nonumber \\
d_W \ \Om &=& [\om, \Om]
\label{eq:A6}
\eea
$i_W(\lambda), \ell_W(\lambda)$ by
\bea
i_W(\lambda) \om &=& \lambda \ \ \ i_W(\lambda)\Om =0 \nonumber \\
\ell_W (\lambda) &=& [i_W (\lambda), d_W]_+ : \nonumber \\
\ell_W(\lambda) \om &=& [\lambda, \om] \nonumber \\
\ell_W (\lambda) \Om &=& [\lambda, \Om]
\label{eq:A7}
\eea

\underline{Definition:} The equivariant cohomolgy of $M$ is the
basic cohomology of $W(\cg) \otimes \Om^\ast(M)$ for the
differential $d_W+d_M$ and the action $i_W(\lambda) + i_M(\lambda),
\ell_W(\lambda) + \ell_M(\lambda)$.

This is the Weil model of equivariant cohomology. 

One can define the
intermediate model according to J.~Kalkman\cite{10} by applying the
algebra automorphism
\be
x \rightarrow e^{-i_M(\om)} x 
\label{eq:A8}
\ee
which transforms the differential into
\be
d_{int} = d_W + d_M + \ell_M(\om) - i_M(\Om)
\label{eq:A9}
\ee
and the operation into
\bea
i_{int} (\lambda) &=& i_W(\lambda) \nonumber \\
\ell_{int} (\lambda) &=& \ell_W (\lambda) + \ell_M (\lambda)
\label{A10}
\eea
From this one easily sees  that the equivariant cohomology is that
of $[ \Om^\ast (M) \otimes S\left( (\mbox{Lie} \ \cg)^\ast\right)
]^{\cg}$ with the differential
\be
d_C = d_M - i_M(\Om)
\label{eq:A11}
\ee
where the superscript $\cg$ denotes $\cg$-invariant elements. This
is the Cartan model \cite{13}, \cite{10}. If $M$ is a principal $\cg$
bundle with a connection $\tilde{\om}$, the mapping
\be
\om \rightarrow \tilde{\om}  \ \ \ \Om \rightarrow \tilde{\Om}
\label{eq:A12}
\ee
where $\tilde{\Om}$ is the curvature of $\tilde{\om}$, maps
isomorphically the equivariant cohomology of $M$ into its basic
cohomology, independently of the choice of $\tilde{\om}$. This is
Cartan's theorem 3 \cite{13}.

There are two standard ways to produce non trivial equivariant
cohomology classes:

i) \cite{12} If the action of $\cg$ can be lifted to a principal
bundle $P(M,K)$ with structure group $K$, and $\Gamma$ is a $\cg$
invariant connection on $P(M,K)$, the intermediate equivariant
curvature is defined as
\be
R^{eq}_{int} \ (\Gamma) = D_{int} \Gamma +
\shalf [\Gamma, \Gamma] =  R(\Gamma) -i_P(\Om) \Gamma
\label{eq:A13}
\ee

One has
\bea
i_{int}(\lambda) \ R^{eq}_{int} (\Gamma) &=&0
\nonumber \\
\ell_{int} (\lambda) \ R^{eq}_{int} &=&
[\lambda,  R^{eq}_{int} (\lambda)]
\label{eq:A14}
\eea

It follows that any $K$ invariant polynomial of Lie $K$,
$P_{inv}$ yields an equivariant "characteristic" cohomology
class. This can be written in the Weil model using Kalkman's
automorphism and is at the root of the construction of topological
observables \cite{6}, \cite{14}.

ii) If $E(X,V)$ is a vector bundle over the manifold $X$, reducible
to $\cg$, one may write
\be
E(X,V) = P(X, \cg) \otimes_\cg V
\label{eq:A15}
\ee
where $P$ is the associated frame bundle.

There is a basic cohomology class, the universal Thom class obtained
as follows \cite{11}:
\be
\tau_0 \equiv \delta(v) \wedge dv =N_0 \int db \ d\bar{\om}
\ e^{i<b,v>+<\bar{\om}, dv>}
\label{eq:A16}
\ee
for some normalization constant $N_0$ where $b$ and $\bar{\om} \in
V^\ast$, the dual of $V, \int d \bar{\om}$ means Berezin
integration, and $<\ ,\ >$ denotes the duality pairing. Introducing
$s$ by
\bea
s \ v &=& dv + \om v \equiv \psi + \om v \nonumber \\
s \ dv &=& -\Om v + \om dv \nonumber \\
s \ \om &=& \Om - \shalf [\om, \om] \nonumber \\
s \  \Om &=& - [\om, \Om] \nonumber \\
s \ \bar{\om} &=& -ib - \bar{\om} \om \nonumber \\
s\ ib &=& -ib\om + \bar{\om} \Om
\label{eq:A17}
\eea

One may write
\be
\tau_0 = \delta(v) (\wedge dv) = N_0 \int db \ d\bar{\om}
\ e^{s<\bar{\om}, V>}
\label{eq:A18}
\ee

It is easy to prove that
\be
\tau = N_0 \int db \ d\bar{\om}
\ e^{s[<\bar{\om}, v> -i(\bar{\om},b)]}
\label{eq:A19}
\ee
where $(\bar{\om},b)$is a $\cg$ invariant bilinear form on
$\cg^\ast$, is an equivariant class of $V$, with fast decrease.
Replacing $\om$ by $\tilde{\om}$, a connection on $P(X,\cg)$, yields
a basic class of $E(X,V)$, once written in the Weil scheme
($\psi_{Weil} = dv - \om v$, whereas $\psi_{int} =
dv$). The extension of the $s$-operation to the integration
variables brings a substantial simplification to the original calculations.

The substitution of $v$ by a section $v(x)$ transforms $\tau$ into
the cohomology class associated with the submanifold of $X$ defined
by $v(x)=0$.

Formula \ref{eq:A19} gives the Matha{\"\i} Quillen representative of
the Thom class of $E(X,V)$ and leads to a gaussianly spread Dirac
current of the submanifold in question.

As a last example, used in the text, let us describe the Ruelle
Sullivan \cite{3},\cite{5} class associated with an invariant closed
form $\om$ on $M$:
\be
\om_{RS} = i(\wedge_\al e_\al) \om
\label{eq:A21}
\ee
where $e_\al$ is a basis of Lie $\cg$.

That $\om_{RS}$ is both closed and invariant follows from the
closedness and invariance of $\om$, and horizontality is trivial
$(i(e_\al) i(e_\al)=0)$.


\begin{thebibliography}{99}
\bibitem{2}The corresponding pages of these notes are available from
the author upon request.
\end{thebibliography}
\end{document}